# Automatic Exam Evaluation based on Brain Computer Interface

Hameda F. Balat
Assistant Lecturer Computer Teacher Preparation Department, Faculty of Specific Education, Damietta University, Egypt

M.A. El-dosuky
Assistant Professor of Computer Science, Faculty of Computers and Information, Mansoura University, Egypt

El-Saeed M. Abd El-Razek
Professor of Using Computer in Education Faculty of Specific Education - Damietta University, Egypt

Magdi Z. Rashed
Professor of Computer Science, Faculty of Computers and Information, Mansoura University, Egypt

## ABSTRACT
Brain computer interface applications can be used to overcome learning problems, especially student anxiety, lack of focus, and lack of attention. This paper introduces a system based on brain computer interface (BCI) to be used in education to measure intended learning outcomes and measure the impact of noise on the degree of system accuracy. This system works online and is based on recorded brain signal dataset. The system can be considered as a special case of P300 speller accepting only letters from A to D. These are the possible answers to multiple-choice questions MCQ. The teacher makes exams, stores them in an exam database and delivers them to students. Students enroll into the system and record their brain signals. Brain signals go through preprocessing phase in which signals undergo low and high pass filter. Then the signals undergo a subsampling and segmentation. The features obtained are used as inputs to Linear Discriminant Analysis (LDA). Gained accuracy is 91%.

## 1. INTRODUCTION
With recent advances in technology, keeping pace with new methods and modern applications is an important issue in education. Brain research, especially the brain computer interface (BCI), is one of the most active areas of research today that based on Electro-Encephalo-Gram (EEG)-based brain activity observation [1].

Brain computer interface applications can be used to overcome learning problems, especially student anxiety, lack of focus, and lack of attention, by reducing student anxiety and increasing learning ability to achieve intended learning outcomes [2].

Electroencephalogram (EEG) is a non-invasive way to scan brain and return its states. But, EEG scans have high rates of noises that make the rendering of EEG signals very difficult. Hence, it is crucial to eliminate these noises in the pre-processing phase before actual analysis of the signals [3].

This paper introduces a system based on brain computer interface (BCI) to be used in education to measure intended learning outcomes and measure the impact of noise on the degree of system accuracy. This system works online and is based on recorded brain dataset.

This paper includes the following sections: section 1 is for introduction, section 2 is for research problem, section 3 is for research objective, section 4 is for research methodology, section 5 is for literature review, section 6 is for proposed system, section 7 is for experiments & results and section 8 is for conclusion & future work.

## 2. RESEARCH PROBLEM
There are many problems that students face while taking the tests, including anxiety, tension, distraction and lack of focus, and this may be due either to the type and method of the test or for other reasons, which negatively lead to the measurement of the intended learning outcomes.

## 3. RESEARCH OBJEVTIVE
This study aims to:

Present a framework for automatic exam that evaluates Intended learning outcomes based on the brain computer interface.

## 4. RESEARCH METHODOLOGY
- The descriptive approach is based on the survey method: (Prepare the theoretical framework and review previous studies related to the study variables).

- Experimental approach :( To experiment proposed system).

## 5. LITERATURE REVIEW
### 5.1 Brain Computer interface (BCI)
BCI is any apparatus capable of translating brain activities into actions. BCI has 4 sub-systems [4]:

- Signal acquisition sub-system for recording brain activities. Recording can be invasive or non-invasive. The most common acquisition is based on Electroencephalography (EEG) [5] [6]. Recently, mobile EEG biosensor-based embedded devices are available [7].

- Signal processing sub-system that discovers significant patterns in brain activities;

- Output apparatus such as a computer screen, robotic arm, or wheelchair; and
- Operating protocol for controlling interaction.

Recently, there is an increasing use of the BCI systems in the field of education, either among students or in teacher-led projects [1].

There are many studies that have used BCI in education, as they have been used in training and reducing math anxiety. To achieve this goal, a research design within subjects with eight waves was used to collect data as a basic methodological way to confirm changes in participants' level of mathematics anxiety through two sessions held on separate days. Analysis of data captured with two training sessions using the BCI Educational Mathematics game showed that mathematics anxiety can be effectively trained and reduced using BCI [8].

BCI was also used in the development, implementation and evaluation of an EEG-based engineering education project, in which students applied theoretical engineering they learned and improved their knowledge and skills in the field of monitoring and evaluation of electrical signals resulting from brain activity and measured by biological sensors. The main objective of the project was to develop and test an interface between the brain and the computer that would be able to measure the average level of interest. The effectiveness of this project-based learning was assessed through student questionnaire questionnaires and analysis of student test results; it was shown that students who participated in the project had higher levels of acquired knowledge [1].

A BCI was also developed to monitor alertness calculated by Think Gear-ASIC Module (TGAM1) and to evaluate outcomes through learning proficiency tests applied in cognitive neuroscience [9].

Interdisciplinary learning tools have been used to expand the types of experiences available to students. This is done by undergraduate students introducing basic BCI concepts using neuroblock which is a visual programming environment that allows users to create applications that rely on real-time neurophysiological data (i.e. brain waves) [10].

A low-cost BCI has been developed, whose characteristics may allow educational institutions to improve the learning and assessment methodologies applied to a particular student [11].

There is a study that explores the possibility of exploiting the information extensively expanded from the human cortex to develop a BCI capable of estimating brain workload and mental efforts during a cognitive task [12].

## 5.2 BCI working mechanism

Figure 1 shows the general BCI architecture, where EEG signals are captured from a device. Then, those signals are pre-processed, extracted, classified, and transmitted to an action output.

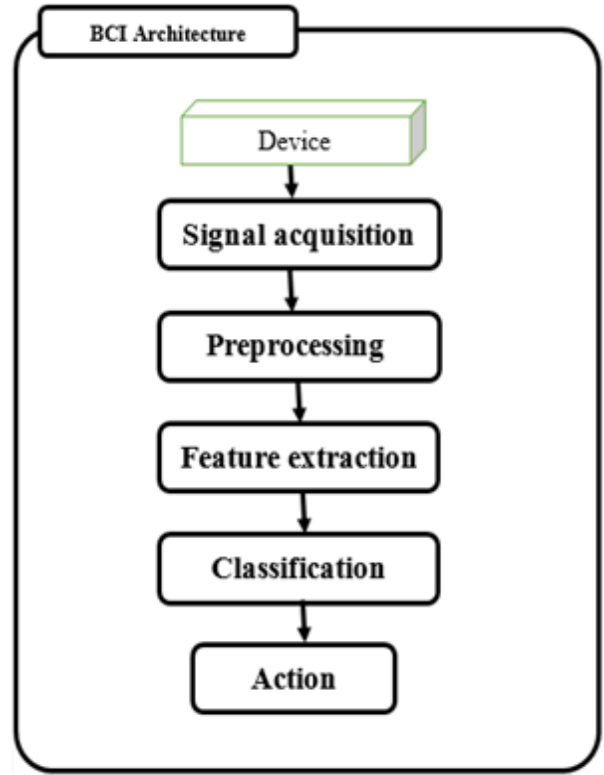

**Fig 1: General BCI Architecture**

During EEG acquisition, electrodes are located at symmetrical skull points, where the conductors can develop as the electrode or poles. In unipolar evolution, the active electrode and the null reference are used, while in the case of the dipole measurement, a possible change is made between two different electrodes.

An internationally acceptable electrode 10/20 electrode arrangement system has been established with respect to the suitability of electrodes, where anatomical attachment points are defined. As in Figure 2 and in order to unambiguously define areas of behavior, each point in the human skull was given a label (Fp, F, T, C, P, O) and a number, where the letters determine the position of the lobe (front pole, frontal, temporal, central, parietal, occipital), and even numbers relate to the hemisphere and odd numbers relate to the left hemisphere [8].

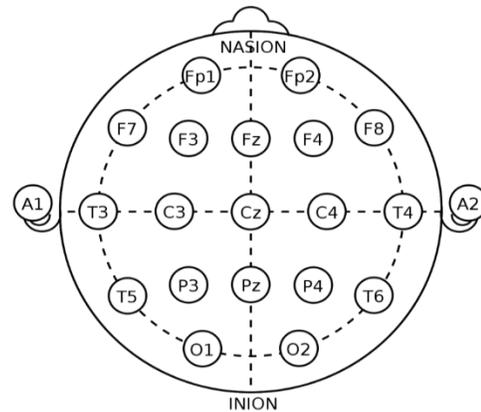

**Fig 2: the international 10/20 system for EEG electrodes [30]**

## 5.3 P300 Speller
The P300 speller is a successful example of BCIs to simulate keyboard input. Its graphical user interface is a grid of alphanumeric characters [13]. P300 speller needs high accuracy and short input times [14].

To achieve the classification phase of P300-speller, Support Vector Machine (SVM) is applied [15]. But, training phase needs large data to ensure classification performance.

In a study, visual mismatch negativity presentation paradigm was compared to a single character presentation paradigm, where the former paradigm has significantly higher classification accuracy [16]. In another study, rapid serial visual presentation (RSVP) is proposed [17].

In another study that uses SVM, the methodology needs insignificant pre-processing and achieves a higher transfer rate [18]. This makes it appropriate for online analysis too.

## 5.4 Noise reduction and removal
One of the problems that face real-time BCI is the difficulty of detecting a target signal due to its noise contamination.

The pollution level increases in recording a non-invasive signal. To detect target signals, the Bayes signal detection can be applied [19].

Many studies showed great interest in removing different noise types. However, there is no optimal noise removal method that does not distort the relevant EEG signal [20].

Extracting relevant EEG signals features is a difficult task. A study suggested using quantitative regression and L1-norm regulation to estimate the EEG spectroscopy [21]. Another study suggested using structured spatial pattern to extract effective features [22].

Another study developed a Linear Discriminant Analysis (LDA) classification method based on four types of features that are Katz Fractal Dimension, Log Variance, Sub band Energy, and Root Mean Square [23].

EEG records are interfered with by various types of noise, which greatly reduces its usefulness. Artificial Neural Networks (ANNs) are an effective in removing EEG interference [24].

## 5.5 Intended learning outcomes
In higher education, learning outcomes are an integral part of the curriculum that corresponds to characteristics of what a student must learn at the end of a course or program [25].

A study aimed to show whether an inverted semester could lead students to increase gains in learning outcomes in Macau, China, and the United States [26].

Another study aimed to determine the effects of a multimedia on educational motivations, improve learning outcomes, efficiency, and class satisfaction [27].

Another study aimed to compare the effectiveness of the flipped classroom inverted learning model on classical lecture-based learning by meta-analysis [28].

## 6. PROPOSED SYSTEM

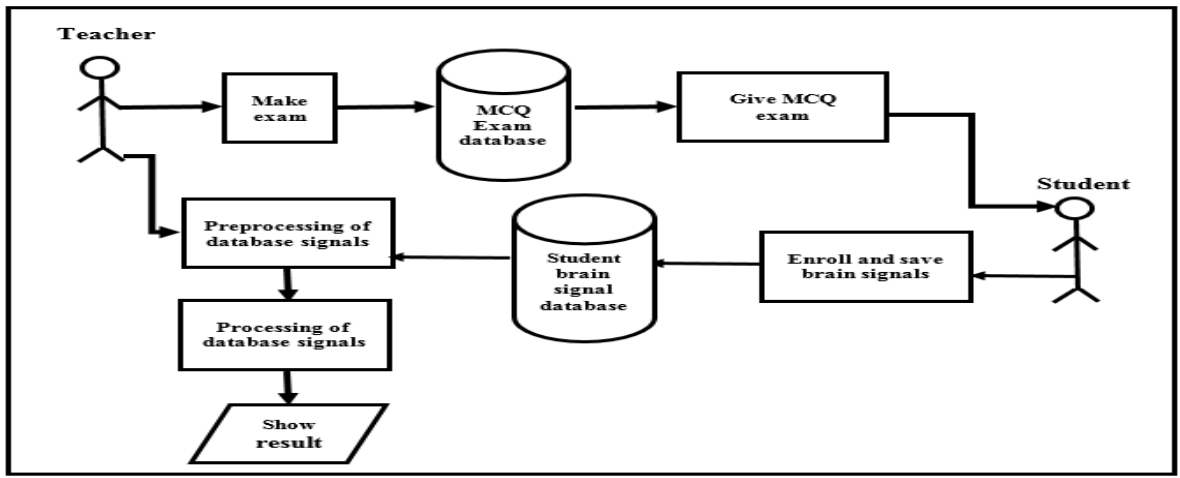

Fig 3: Proposed Architecture

The system can be considered as a special case of P300 speller accepting only letters from A to D. These are the possible answers to multiple-choice questions MCQ.

The teacher makes exams, stores them in an exam database and delivers them to students. Students enroll into the system and record their brain signals.

Brain signals go through preprocessing phase in which signals undergo low and high pass filter, as shown in figure 4. Then the signals undergo a subsampling and segmentation as shown in figure 5.

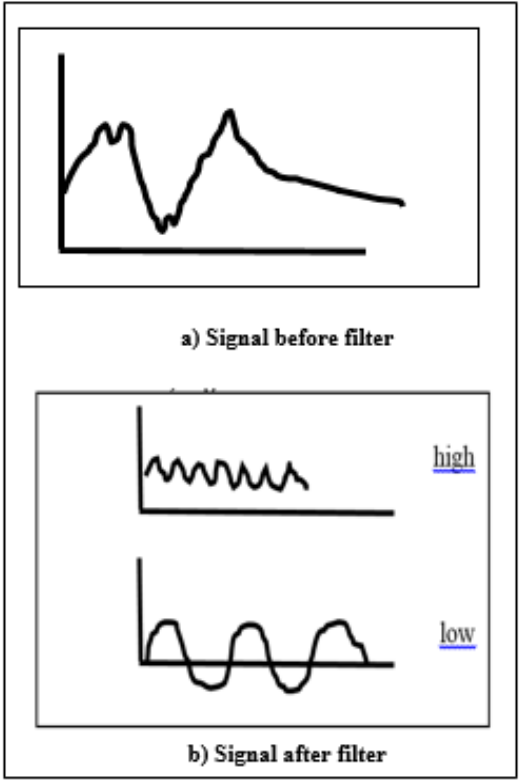

**Fig 4: high and low pass filtering**

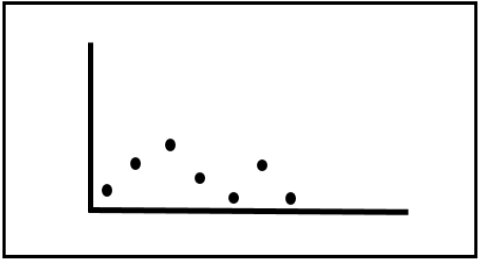

**Fig 5: sub sampling**

The features obtained are used as inputs to Linear Discriminant Analysis ( LDA).

LDA commences with finding the mean signal of the entire sample signals using:

$$\mu_i = \frac{1}{N}\sum_{k=1}^{N} x_k \quad (1)$$

Where μi is the mean vector of samples belonging to class Xi, Ni is the total number of training samples in class Xi, and Xk represents the samples belonging to class Xi. If μ is the mean of the sample space and C is the number of distinct classes, then the within-scatter matrix (SW) and the between-scatter matrix (SB) of the data are measured, using:

$$S_W = \sum_{i=1}^{C}\sum_{x_k \in X_i}(x_k - \mu_i)(x_k - \mu_i)^T \quad (2)$$

$$S_B = \sum_{i=1}^{C}(\mu_i - \mu)(\mu_i - \mu)^T \quad (3)$$

Then LDA selects optimal W in a way that the ratio of the between-class scatter and the within-class scatter is maximized:

$$W_{opt} = argmax \frac{W^T S_B W}{W^T S_W W} \quad (4)$$

## 7. EXPERIMENTS & RESULTS

The system is tested on many students of the second year, Computer Teacher Preparation Department, Faculty of Specific Education, Damietta University. Figure 6 shows two of volunteering students. Figure 7 shows the testing environment.

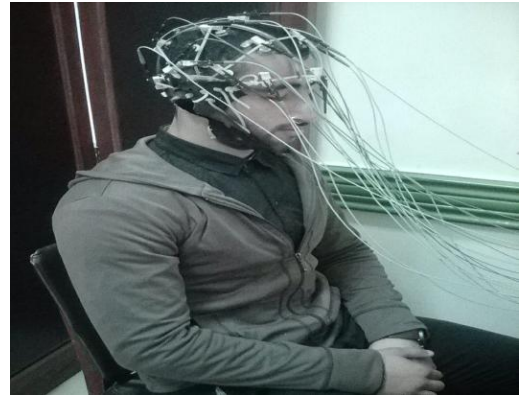

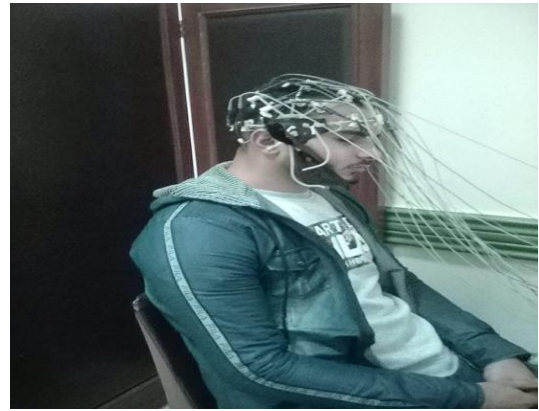

**Fig 6: two of volunteering students**

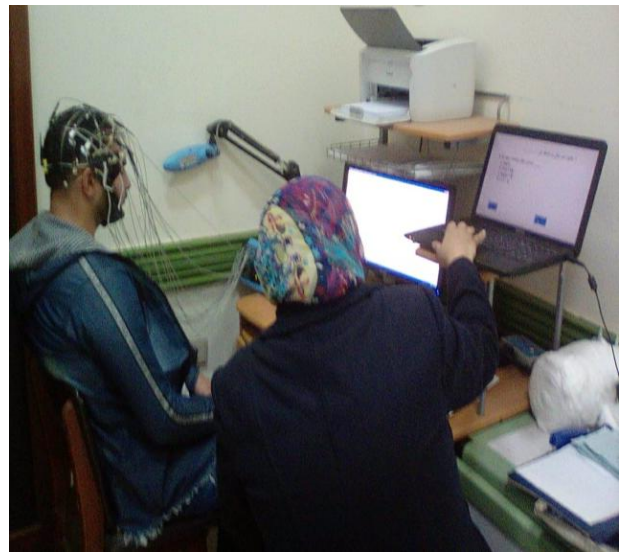

**Fig 7: Testing Environment**

The gained accuracy is 91%. To study the effect of noise on accuracy, we gradually add noise and recognizing its effect on accuracy. Initial accuracy is 91%. It reaches zero accuracy at 98% noise, as shown in table 1and Figure 8.

**Table 1. Effect of noise on accuracy**

| n. | Error | Accuracy Before | Accuracy After |
|---|---|---|---|
| 1 | 1 | 91 | 91 |
| 2 | 2 | 91 | 91 |
| 3 | 3 | 91 | 91 |
| 4 | 4 | 91 | 91 |
| 5 | 5 | 91 | 91 |
| 6 | 6 | 91 | 91 |
| 7 | 7 | 91 | 91 |
| 8 | 8 | 91 | 91 |
| 9 | 9 | 91 | 91 |
| 10 | 10 | 91 | 90 |
| 11 | 11 | 91 | 90 |
| 12 | 12 | 91 | 89 |
| 13 | 13 | 91 | 88 |
| 14 | 14 | 91 | 87 |
| 15 | 15 | 91 | 86 |
| 16 | 16 | 91 | 85 |
| 17 | 17 | 91 | 84 |
| 18 | 18 | 91 | 83 |
| 19 | 19 | 91 | 80 |
| 20 | 20 | 91 | 80 |
| 21 | 21 | 91 | 79 |
| 22 | 22 | 91 | 78 |
| 23 | 23 | 91 | 77 |
| 24 | 24 | 91 | 76 |
| n. | Error | Accuracy Before | Accuracy After |
| 25 | 25 | 91 | 75 |
| 26 | 26 | 91 | 75 |
| 27 | 27 | 91 | 75 |
| 28 | 28 | 91 | 74 |
| 29 | 29 | 91 | 73 |
| 30 | 30 | 91 | 72 |
| 31 | 31 | 91 | 71 |
| 32 | 32 | 91 | 70 |
| 33 | 33 | 91 | 69 |
| 34 | 34 | 91 | 68 |
| 35 | 35 | 91 | 68 |
| 36 | 36 | 91 | 68 |
| 37 | 37 | 91 | 67 |
| 38 | 38 | 91 | 66 |
| 39 | 39 | 91 | 65 |
| 40 | 40 | 91 | 64 |
| 41 | 41 | 91 | 63 |
| 42 | 42 | 91 | 62 |
| 43 | 43 | 91 | 61 |
| 44 | 44 | 91 | 60 |
| 45 | 45 | 91 | 59 |
| 46 | 46 | 91 | 58 |
| 47 | 47 | 91 | 57 |
| 48 | 48 | 91 | 56 |
| 49 | 49 | 91 | 55 |
| 50 | 50 | 91 | 54 |
| 51 | 51 | 91 | 53 |
| 52 | 52 | 91 | 52 |
| 53 | 53 | 91 | 51 |
| 54 | 54 | 91 | 50 |
| 55 | 55 | 91 | 50 |
| 56 | 56 | 91 | 50 |
| 57 | 57 | 91 | 50 |
| 58 | 58 | 91 | 49 |
| 59 | 59 | 91 | 48 |
| 60 | 60 | 91 | 47 |
| 61 | 61 | 91 | 46 |
| 62 | 62 | 91 | 45 |
| 63 | 63 | 91 | 44 |
| 64 | 64 | 91 | 43 |
| 65 | 65 | 91 | 42 |
| 66 | 66 | 91 | 42 |
| 67 | 67 | 91 | 42 |
| 68 | 68 | 91 | 41 |
| 69 | 69 | 91 | 40 |
| 70 | 70 | 91 | 39 |
| 71 | 71 | 91 | 38 |
| 72 | 72 | 91 | 37 |
| 73 | 73 | 91 | 36 |
| 74 | 74 | 91 | 35 |
| 75 | 75 | 91 | 34 |
| 76 | 76 | 91 | 33 |
| 77 | 77 | 91 | 32 |
| 78 | 78 | 91 | 30 |
| 79 | 79 | 91 | 29 |
| 80 | 80 | 91 | 27 |
| 81 | 81 | 91 | 25 |
| 82 | 82 | 91 | 23 |

**Table 2. Effect of noise on accuracy**

| n. | Error | Accuracy Before | Accuracy After |
|---|---|---|---|
| 83 | 83 | 91 | 21 |
| 84 | 84 | 91 | 19 |
| 85 | 85 | 91 | 17 |
| 86 | 86 | 91 | 15 |
| 87 | 87 | 91 | 13 |
| 88 | 88 | 91 | 11 |
| 89 | 89 | 91 | 9 |
| 90 | 90 | 91 | 7 |
| 91 | 91 | 91 | 5 |
| 92 | 92 | 91 | 4 |
| 93 | 93 | 91 | 3 |
| 94 | 94 | 91 | 3 |
| 95 | 95 | 91 | 2 |
| 96 | 96 | 91 | 2 |
| 97 | 97 | 91 | 1 |
| 98 | 98 | 91 | 1 |
| 99 | 99 | 91 | 0 |
| 100 | 100 | 91 | 0 |

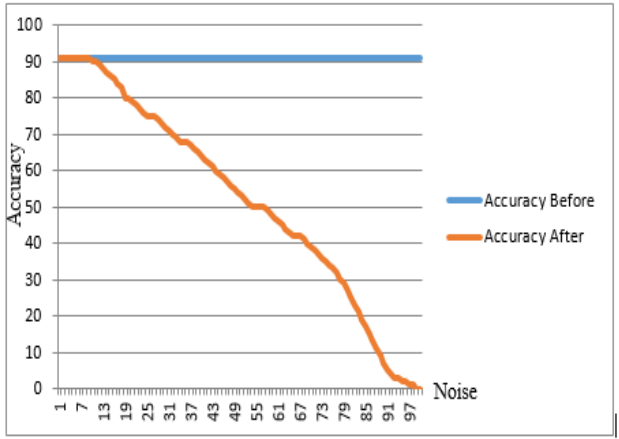

**Fig 8: Relationship between Noise and Accuracy**

The spectrum is decomposed from violete to red colors. Alpha signal is represeted as red in EEG while other signals are at the violet direction.

Alpha is considered the attention signal. While the other signals are considered noise.

## 8. CONCLUSION & FUTURE WORK

The proposed system can be considered as a special case of P300 speller accepting only letters from A to D. These are the possible answers to multiple-choice questions MCQ. The teacher makes exams, stores them in an exam database and delivers them to students. Students enroll into the system and record their brain signals. Brain signals go through preprocessing phase in which signals undergo low and high pass filter. Then the signals undergo a subsampling and segmentation. The features obtained are used as inputs to Linear Discriminant Analysis ( LDA). Gained accuracy is 91%.

Future directions may consider other type of textual exam questions other than MCQ. Another future direction may consider using specialized hardware such as emotive headset [31].